\documentclass[twocolumn]{aastex62}

\usepackage{color}

\graphicspath{{./}{figures/}}

\received{January 1, 2018}
\revised{January 7, 2018}
\accepted{\today}
\submitjournal{ApJ}

\shorttitle{Diffuse radio emission in the Phoenix cluster}
\shortauthors{Raja et al.}

\begin{document}

\title{PROBING THE ORIGIN OF DIFFUSE RADIO EMISSION IN THE COOL-CORE OF THE PHOENIX GALAXY CLUSTER}

\correspondingauthor{Ramij Raja}
\email{phd1601121008@iiti.ac.in}

\author[0000-0001-8721-9897]{Ramij Raja}
\affil{Discipline of Astronomy, Astrophysics
and Space Engineering, Indian Institute of Technology Indore, Simrol, 453552, India}
\author[0000-0002-1372-6017]{Majidul Rahaman}
\affil{Discipline of Astronomy, Astrophysics
and Space Engineering, Indian Institute of Technology Indore, Simrol, 453552, India}
\author[0000-0002-5333-1095]{Abhirup Datta}
\affil{Discipline of Astronomy, Astrophysics
and Space Engineering, Indian Institute of Technology Indore, Simrol, 453552, India}
\author[0000-0002-4468-2117]{Jack O. Burns}
\author[0000-0003-0025-6762]{Brian Alden}
\affil{Center for Astrophysics \& Space Astronomy, Department of Astrophysical \& Planetary Sciences, 389 UCB, University of Colorado,
Boulder, CO 80309, USA}
\author[0000-0002-5880-2730]{H. T. Intema}
\affil{Leiden Observatory, Leiden University, PO Box 9513, 2300 RA Leiden, The Netherlands}
\affil{International Centre for Radio Astronomy Research, Curtin University, GPO Box U1987, Perth, WA 6845, Australia}
\author[0000-0002-0587-1660]{R. J. van Weeren}
\affil{Leiden Observatory, Leiden University, PO Box 9513, 2300 RA Leiden, The Netherlands}
\author[0000-0002-2703-0790]{Eric J. Hallman}
\affil{Center for Astrophysics \& Space Astronomy, Department of Astrophysical \& Planetary Sciences, 389 UCB, University of Colorado,
Boulder, CO 80309, USA}
\author[0000-0003-2196-6675]{David Rapetti}
\affil{Center for Astrophysics \& Space Astronomy, Department of Astrophysical \& Planetary Sciences, 389 UCB, University of Colorado,
Boulder, CO 80309, USA}
\affiliation{NASA Ames Research Center, Moffett Field, CA 94035, USA}
\author[0000-0003-4046-6959]{Surajit Paul}
\affil{Department of Physics, Savitribai Phule Pune University, Pune 411007, India}



\begin{abstract}
\noindent
Cool core galaxy clusters are considered to be dynamically relaxed clusters with regular morphology and highly X-ray luminous central region. However, cool core clusters can also be sites for merging events that exhibit cold fronts in X-ray and mini-halos in radio. We present recent radio/X-ray observations of the Phoenix Cluster or SPT-CL J2344-4243 at the redshift of $z=0.596$. Using archival {\it Chandra} X-ray observations, we detect spiraling cool gas around the cluster core as well as discover two cold fronts near the core. It is perhaps the most distant galaxy cluster to date known to host cold fronts. Also, we present JVLA\footnote{Jansky Very Large Array\\ \url{https://science.nrao.edu/facilities/vla}} 1.52 GHz observations of the minihalo,  previously discovered at 610 MHz with GMRT\footnote{Giant Metrewave Radio Telescope\\ \url{http://www.gmrt.ncra.tifr.res.in}} observations in the center of the Phoenix galaxy cluster.
The minihalo flux density at 1.52 GHz is $9.65 \pm 0.97$ mJy with the spectral index between 610 MHz and 1.52 GHz being $-0.98 \pm 0.16$\footnote{$S_{\nu} = \nu^{\alpha}$ where $S_{\nu}$}.
A possible origin of these radio sources is turbulence induced by sloshing of the gas in the cluster core. 

\end{abstract}



\section{Introduction} \label{sec:intro}

Recent studies have firmly established our understanding of the intracluster medium (ICM) as formed by hot diffuse plasma, relativistic particles, and large scale magnetic fields (for reviews, see \citealt{Feretti_rev,Brunetti_rev, vanWeeren2019}). 
The observed diffuse radio emission in the galaxy clusters are typically three types: (1) Giant Radio Halos (GRHs), (2) Relics \& Radio Phoenices, and (3) Minihalos (MHs) (e.g.~\citealt{Feretti_rev}). The GRHs are $\sim$Mpc scale radio emission found in merging clusters. They are roundish and located in the central region of the cluster. The radio relics, on the other hand, are found in the periphery of merging clusters. They are of elongated morphology with a typical size of about $0.5-2$ Mpc. The radio phoenices are located close to the cluster center compared to the relics. They are found in both merging and relaxed clusters, having a roundish, elongated, or filamentary morphology with a typical size of $\lesssim0.3-0.4$ Mpc \citep{vanWeeren2019}.
Minihalos are $\sim100-500$ kpc scale diffuse radio sources of synchrotron origin with roundish morphology, found only surrounding the central Brightest Cluster Galaxy (BCG) in some relaxed, cool-core clusters. 
According to \citet{Giacintucci2017}, diffuse radio emission is categorized as a \lq\lq minihalo\rq\rq\ if (1) it does not resemble radio lobes, tails nor has any connection with the central AGN, (2) minihalo radius is $> 50$ kpc and (3) a maximum radius of $0.2R_{500}$. This $r \lesssim 0.2R_{500}$ radius separates the outer self-similarly evolving ICM region and the inner region where cooling, stellar feedback, AGN feedback, sloshing, etc., becomes more important \citep{McDonald2017}.
Minihalos found at the center of the Perseus cluster represent the prototypes of this class of radio sources (e.g., \citealt{Pedlar1990,Burns1992,Sijbring1993,Gendron2017}). A recent study by \citet{Giacintucci2017} revealed that minihalos are not that rare as was previously believed, and are very common ($\sim 80\%$) in massive ($>6\times 10^{14} \mathrm{M_{\odot}}$) cool core clusters.

{\it Chandra} with its high spatial and spectral resolution X-ray observations, has revealed many remarkable physical processes that are happening in galaxy clusters.
The minor merging events of sub-clusters can perturb the cluster central gravitational potential, resulting in gas \lq\lq sloshing\rq\rq\ in the cluster core. The presence of these phenomena is detected from the temperature jumps and corresponding density decrease across the interface of moving cold gas in the cluster core region. These are \lq\lq cold fronts\rq\rq\ or contact discontinuities ubiquitous in cool core clusters as the only necessary condition being a steep radial entropy profile (\citealt{Ascasibar2006}).
Cold fronts are more subtle in density profiles of cool core clusters than of those found in mergers, and they occur closer to the cluster center ($r \lesssim 100$ kpc)~\citep{Markevitch2007}.

The origin of the minihalo emission in the galaxy clusters is of great interest still today.
The radiative lifetime of relativistic electrons is much shorter than their diffusion timescale, prohibiting the formation of diffuse radio sources of the order of the minihalo radius only by diffusion. Thus, some other mechanism is required to explain their origin. \citet{Burns1992} proposed that the absence of Mpc scale radio emission in the Perseus cluster was because of the lack of magnetic fields in the outer region of the cluster core due to the cooling flow suppression of turbulence amplified magnetic fields. Later, \citet{Gitti2002,Gitti2004} suggested that in the presence of magnetic fields, re-acceleration of {\it in situ} electrons by turbulence caused by the cooling flow can produce radio structures of this scale.  
However, signatures of the classical \lq\lq cooling flow\rq\rq\ model were not observed
in recent observations \citep{Peterson2006}, whereas large-scale sloshing gas has been observed quite frequently in X-ray observations (for reviews, see \citealt{Markevitch2007,Bohringer2010}).
\citet{Fujita2004} suggested that sloshing can generate significant turbulence in the core of clusters, and \citet{Mazzotta2008} found a striking spatial correlation between minihalo emission and cold fronts possibly arising from gas sloshing. They found that the diffuse radio emission was contained within the cold fronts, and later on, hydrodynamic simulations from \citet{ZuHone2013} reproduced this finding. Such a feature has been seen in other clusters as well~\cite[see e.g.,][]{Giacintucci2014}.
Alternatively, in the hadronic model, inelastic collisions between thermal and relativistic cosmic ray protons (CRp) may provide the continuous supply of relativistic electrons needed for large scale synchrotron emission (e.g., \citealt{Pfrommer2004,Fujita2007,Keshet2010,Keshet2010a}).
Recently, another possible scenario was put forward by \citet{Omar2019} where multiple supernovae type Ia (SNIa) events over the 100 Myr period can provide synchrotron emitting relativistic electrons on about 500 kpc scale.

In this paper, we present radio and X-ray analyses of the Phoenix cluster or SPT-CL J2344-4243 \citep{Williamson2011} as a multi-wavelength study of the diffuse radio emission (radio minihalo) surrounding the central radio galaxy and to understand its origin.
A short review of previous observational work carried out on the Phoenix cluster is reported in Section~\ref{sec:review}. The radio and X-ray observations we employ in this paper, as well as the corresponding data reductions, are described in Section~\ref{sec:observation data reduction}. The results from our radio and X-ray data analyses are reported in Section~\ref{sec:radio_emission} and Section~\ref{sec:xray results}, respectively. 
Finally, a summary of the study and our conclusions are presented in Section~\ref{sec:conclude}.

Here we adopt a $\Lambda$CDM cosmology with $H_0 = 70$ km s$^{-1}$ Mpc$^{-1}$, $\Omega_{\mathrm{m}} = 0.3$ and $\Omega_\Lambda = 0.7$. At the cluster redshift $z = 0.596$, $1\arcsec$ corresponds to a physical scale of 6.664 kpc.


\begin{deluxetable*}{lcccccc}[!t]
\tablecaption{Radio Observations of the Phoenix cluster \label{tab:radio_obs}}
\tablecolumns{7}
\tablewidth{0pt}
\tablehead{
\colhead{Array} & \colhead{Project} & \colhead{Frequency} & \colhead{Bandwidth (MHz)} & \colhead{Obs. date} & \colhead{Obs. time (hrs)} & \colhead{PI}}
\startdata
JVLA-CnB & 14B-397 & $1-2$ GHz & 1024 & 23-01-2015 & 1.67 & A. Datta\\
GMRT & 24\_007 & 610 MHz & 32 & 14 \& 15-06-2013 & 10 & R. J. van Weeren\\
\enddata
\end{deluxetable*}

\begin{deluxetable}{lc}[!b]
\tablecaption{Cluster and Minihalo Properties \label{tab:info}}
\tablecolumns{2}
\tablewidth{0pt}
\tablehead{Parameter & Value}
\startdata
RA DEC & 23h44m42.2s -42d43m08s\\
$z$ & $0.596 \pm 0.002$ \\
$R_{500}$ (Mpc) & 1.3\\
$M_{500}$ ($10^{14}\ \mathrm{M_{\odot}}$) & $12.6^{+2.0}_{-1.5}$\\
$L_{X_{2-10 \mathrm{keV}},500}$ ($10^{44}$ erg s$^{-1}$) & $82^{+1}_{-2}$\\
$T_{500}$ (keV) & $13.0^{+2.4}_{-3.4}$\\
$\dot{M}_{\mathrm{cool}}$ ($\mathrm{M_{\odot}}$ yr$^{-1}$) & $2366 \pm 60$\\
$t_{\mathrm{cool},0}$ (Gyr) & $0.18^{+0.01}_{-0.02}$\\
$K_0$ (keV cm$^2$) & $16^{+2}_{-3}$\\
\hline
$S^{\mathrm{MH}}_{\ 610\ \mathrm{MHz}}$ (mJy) & $22.54 \pm 2.26$\\
$S^{\mathrm{MH}}_{\ 1.52\ \mathrm{GHz}}$ (mJy) & $9.65 \pm 0.97$\\
$R^{\mathrm{MH}}$ (Mpc) & $0.31$\\
$\alpha^{1520}_{610}$ & $-0.98 \pm 0.16$\\
$P^{\mathrm{MH}}_{\ 1.4\ \mathrm{GHz}}$ (10$^{24}$ W Hz$^{-1}$) & $14.38 \pm 1.80$\\
\enddata
\tablecomments{Cluster properties are reported from \citealt{McDonald2012,McDonald2019}.}
\end{deluxetable}

\section{The Phoenix cluster}
\label{sec:review}
As reported by \citet{Williamson2011}, the Phoenix cluster was first detected via the Sunyaev--Zel'dovich effect using the South Pole Telescope (SPT).
Follow up detections of the same were also reported by \citet{Bleem2015,Planck2016}.
Since its discovery, multiple X-ray and radio studies have been carried out on this cluster. 
The Phoenix cluster is massive ($M_{500}$ $\approx$ $12.6 \times 10^{14}$ $h_{70}^{-1}$ $\mathrm{M_{\odot}}$), situated at a redshift of $z = 0.596$ and is the most X-ray luminous ($L_{2-10\ \mathrm{keV}} = 8.2 \times 10^{45}$ erg s$^{-1}$) cluster known till date \citep{McDonald2012,McDonald2019}. It is also the most extreme cool core cluster known, with a steep entropy profile, extreme star formation rate (SFR = $798 \pm 42$ $\mathrm{M_{\odot}}$ yr$^{-1}$; \citealt{McDonald2013}) and the mass deposition rate being $\dot{M}_{\mathrm{cool}} = 2366 \pm 60$ $\mathrm{M_{\odot}}$ yr$^{-1}$, perhaps the rare example of hosting runaway ICM cooling \citep{McDonald2019,McDonald2019a}. The relaxation criteria of \citet{Nurgaliev2017} suggests that the cluster has not experienced any major merger event at least in the last 3 Gyr \citep{McDonald2019}.
The global cluster properties are presented in Table \ref{tab:info}.

\citet{vanWeeren2014} discovered the presence of diffuse radio emission in the Phoenix cluster with 610 MHz observations, surrounding the central BCG with no obvious connection to the radio galaxies and classified the source as a radio minihalo.
\citet{Hlavacek2015} first discovered X-ray cavities in the cluster core resulted from the AGN feedback.
\citet{McDonald2015} found a connection between the X-ray cavities and radio emission from the BCG with 610 MHz GMRT observations where the radio emission appears to be oriented towards the X-ray cavities.
Recently, with the VLA 8-12 GHz observations of the Phoenix cluster, \citet{McDonald2019a} reported the near correspondence of these X-ray cavities with the radio jets revealing the presence of mechanical feedback from the central AGN.

In the next section, we present new the JVLA $1-2$ GHz observation using the CnB configuration as well as archival Chandra data on the cluster.

\begin{figure*}[!t]
\centering
\begin{tabular}{cc}

\includegraphics[width=\columnwidth]{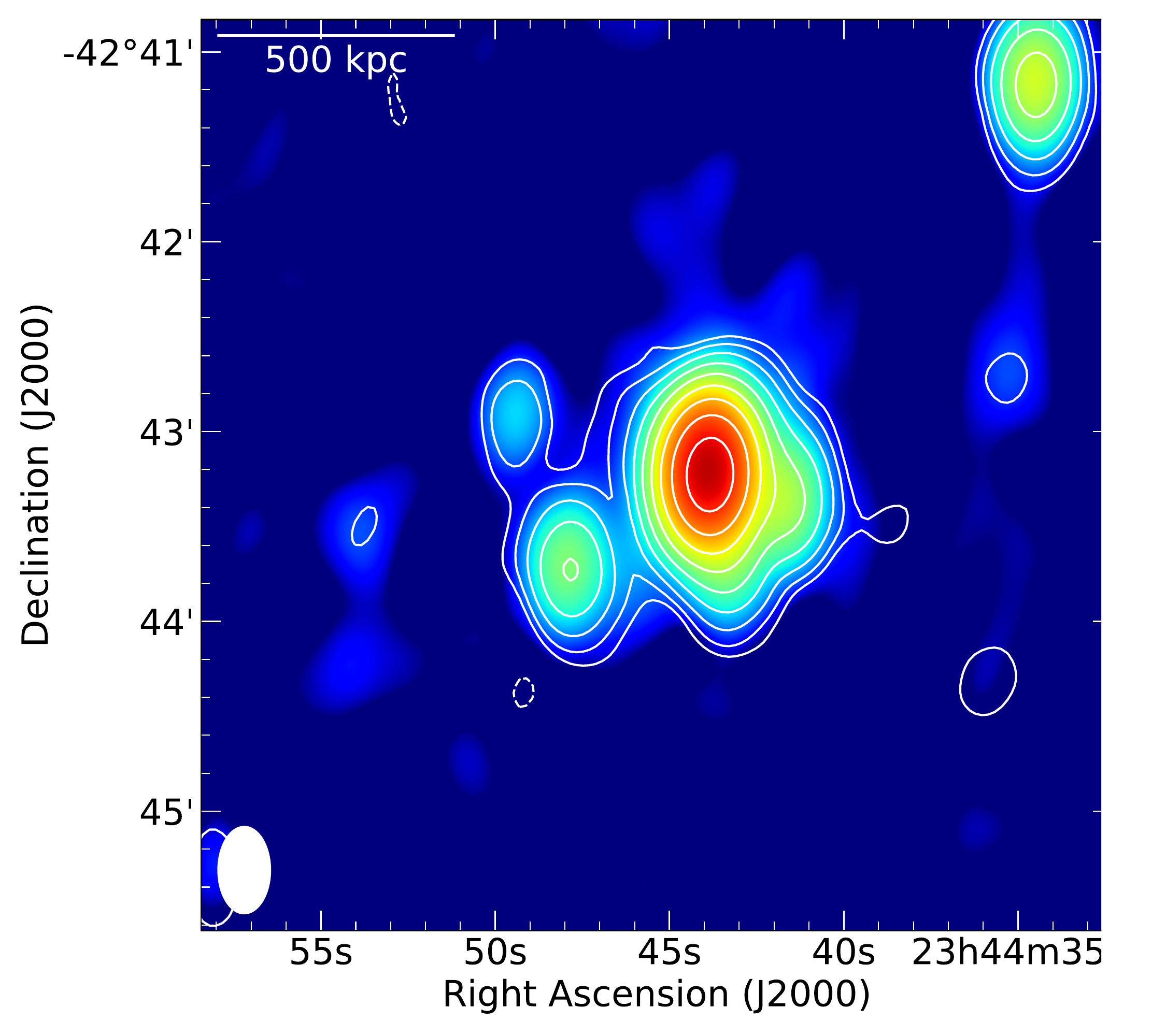} &
\includegraphics[width=\columnwidth]{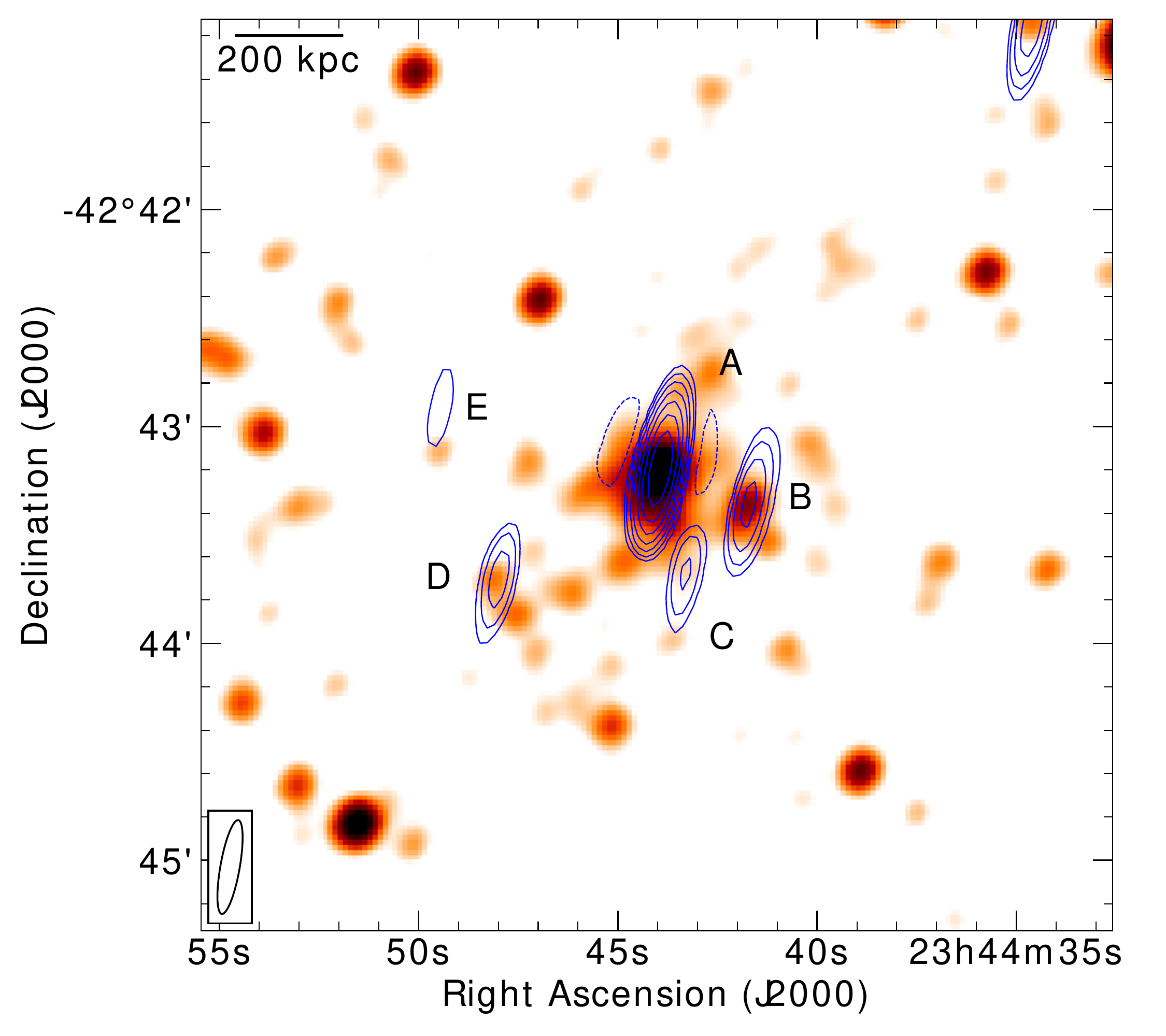}
\end{tabular}
\caption{{\it Left:} JVLA 1.52 GHz low resolution image overlaid with GMRT 610 MHz radio contours.
The radio contours are drawn at levels $[-1,1,2,4,8,...]\times 4\sigma_{\mathrm{rms}}$ where $\sigma_{\mathrm{rms}}$= 50 $\mu$Jy beam$^{-1}$. The beam size for both 1.5 GHz and 610 MHz images is $28\arcsec \times 17\arcsec$  and is shown in the bottom left corner. The map noise of the 1.52 GHz image is $\sigma_{\mathrm{rms}}$= 25 $\mu$Jy beam$^{-1}$.
{\it Right:} {\it WISE} $3.4\ \mu$m infrared image overlaid with JVLA 1.52 GHz high resolution ($26.3\arcsec \times 5.3\arcsec$) image contours (blue). The contours are drawn at the same levels as the previous but with $\sigma_{\mathrm{rms}}$= 60 $\mu$Jy beam$^{-1}$.
Negative contours are indicated with dotted lines.
}
\label{fig:radio_image-cont}
\end{figure*}

\section{observations and data analysis} \label{sec:observation data reduction}

\subsection{JVLA L-Band Observations}
Observation of the Phoenix cluster was carried out with JVLA CnB configuration on 23rd January 2015, for a total of 1.67 hrs of observing time. It was observed in $1-2$ GHz band in full polarization mode with a total bandwidth of 1 GHz.
The 1 GHz observing bandwidth was divided into 16 spectral windows, and each spectral window was further divided into 64 channels.

The CASA (Common Astronomy Software Applications) package developed by NRAO\footnote{National Radio Astronomy Observatory} was used for data reduction purposes. 
We used the VLA calibration pipeline\footnote{\url{https://science.nrao.edu/facilities/vla/data-processing/pipeline}} to perform the basic flagging and calibration using CASA~\citep{McMullin2007}. The pipeline iteratively performs calibration and bad data flagging. From the output calibrated data, usable spectral windows were separated using CASA task {\it SPLIT}. To remove any remaining bad data, careful inspection and flagging were done manually. Around 40\% of on source data was flagged in the process. Calibrator 3C48 was used for bandpass and phase calibration. The flux density of the calibrator was set according to \citet{Perley2013}. Imaging was done using the CASA task {\it CLEAN}. MSMFS \citep{Rau2011} imaging was performed using 2 Taylor terms for spectral modeling. For wide-field imaging, 863 w-projection planes were used. The Briggs weighting \citep{Briggs1995} scheme was used for imaging, as specifically pointed out below. Few rounds of phase-only self-calibration were performed to remove residual phase variations. For self-calibration steps, we set Briggs robust parameter to -1.

\subsection{GMRT 610 MHz Observations}
We have re-analyzed the GMRT 610 MHz archival data of the Phoenix cluster to estimate the spectral index of the minihalo emission. The observations were performed with GMRT on 14 \& 15 June, 2013 in dual-polarization mode with a total observing time of about 10 hr.
This is Legacy Data that used the GMRT Software Backend (GSB) correlator with a total of 32 MHz of bandwidth divided into 256 channels.

Data reduction was done using SPAM\footnote{\url{http://www.intema.nl/doku.php?id=huibintemaspam}} (Source Peeling and Atmospheric Modeling) \citep{Intema2009,Intema2017}, an automated Python-based pipeline employing the Astronomical Image Processing System (AIPS) to reduce high resolution low frequency radio interferometric data. 
The data reduction starts with automatic RFI excision, bandpass, and gain calibration. 3C48 was used for flux calibration, and the flux density was set according to the \citet{Scaife2012} scale. Several rounds of self-calibration were performed, followed by direction dependent calibration of the bright sources. A detailed description of the working pipeline can be found in \citet{Intema2009,Intema2017}. Imaging was performed on the final calibrated data using the CASA task {\it CLEAN}.

\subsection{Chandra X-ray Observations} \label{subsec:xray data analysis}

We used all {\it Chandra} archival data available at the time of data analysis with observation IDs 13401, 16135, 16545, 19581, 19582, 19583, 20630, 20631, 20634, 20635, 20636, 20797. All the observations were made with the ACIS-I instrument in VFAINT mode. Combining all 12 observations, a total of about $\sim$538 ks of clean exposure was obtained.


Our group has developed a semi-automated pipeline, which was initially written in IDL and bash-script \citep{Datta2014,Schenck2014}. Recently, it is developed into a more automated Python-based pipeline called ClusterPyXT\footnote{\url{https://github.com/bcalden/ClusterPyXT}}~\citep{Alden2019}. 

However, for the analysis in this paper, we have used the older version of the pipeline only \citep{Datta2014,Schenck2014}. This pipeline uses \textit{Chandra Interactive Analysis of Observations (CIAO)\footnote{\url{http://cxc.harvard.edu/ciao/}}} along with \textit{Chandra Calibration Database (CALDB)} and \textit{X-Ray Spectral Fitting Package (XSPEC\footnote{\url{https://heasarc.gsfc.nasa.gov/xanadu/xspec/}}}; \citealt{Arnaud1996}). Here, data analysis was performed using CIAO-4.9 and calibrated with CALDB-4.7.9, the most recent version available during the analysis. Once the cluster name and ObsIds are provided, it automatically downloads the data using CIAO, cleans it, and merges all ObsIds together.

\begin{figure}
\centering
\includegraphics[width=\columnwidth]{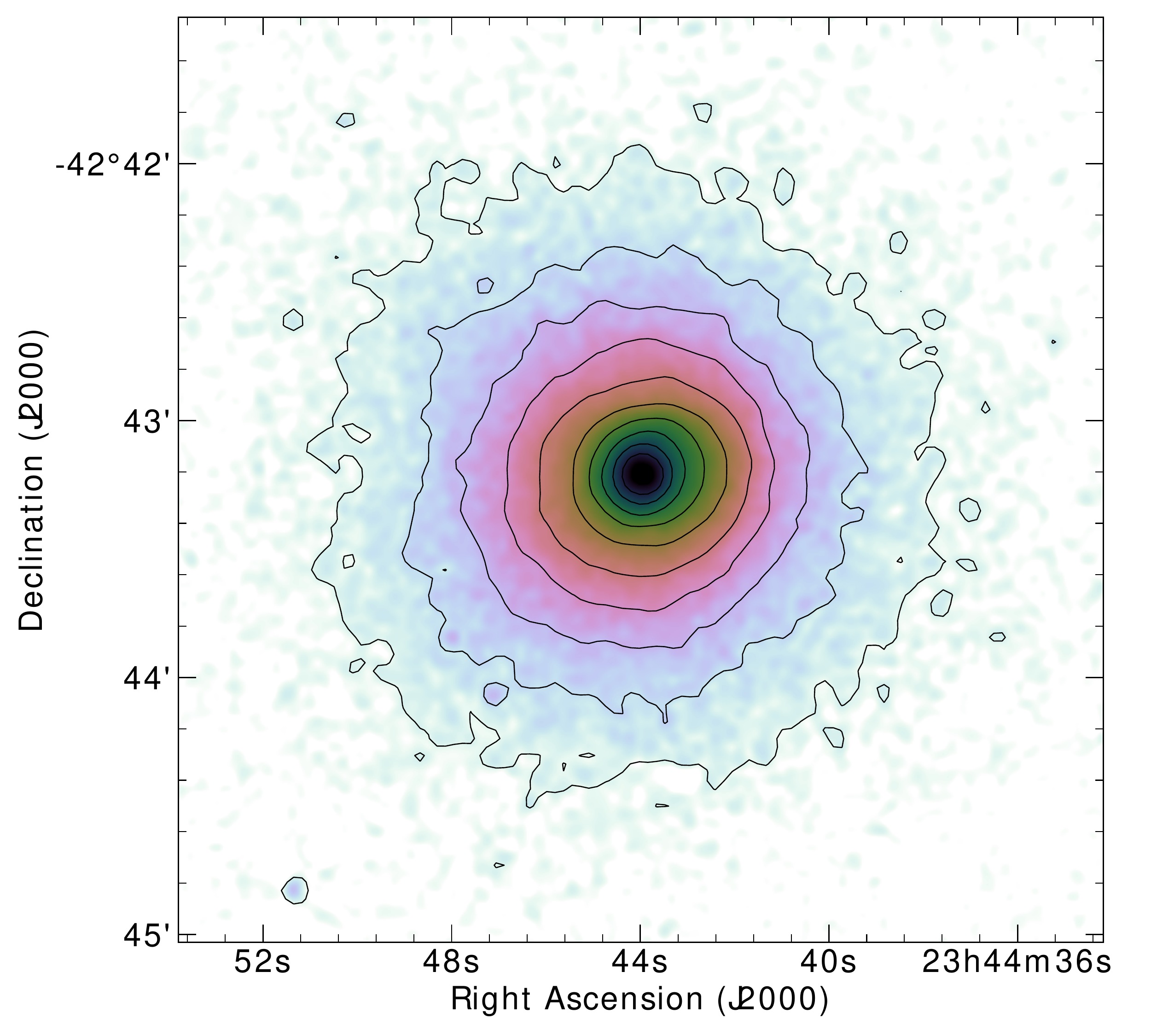} 
\caption{$\it Chandra$ ACIS-I point source subtracted image of the Phoenix cluster in the $0.7-8.0$ keV band overlaid with X-ray contours. The contour levels are spaced by a factor of 2.}
\label{fig:xray}
\end{figure}

\begin{figure*}[!t]
\centering
\begin{tabular}{cc}
\includegraphics[width=\columnwidth]{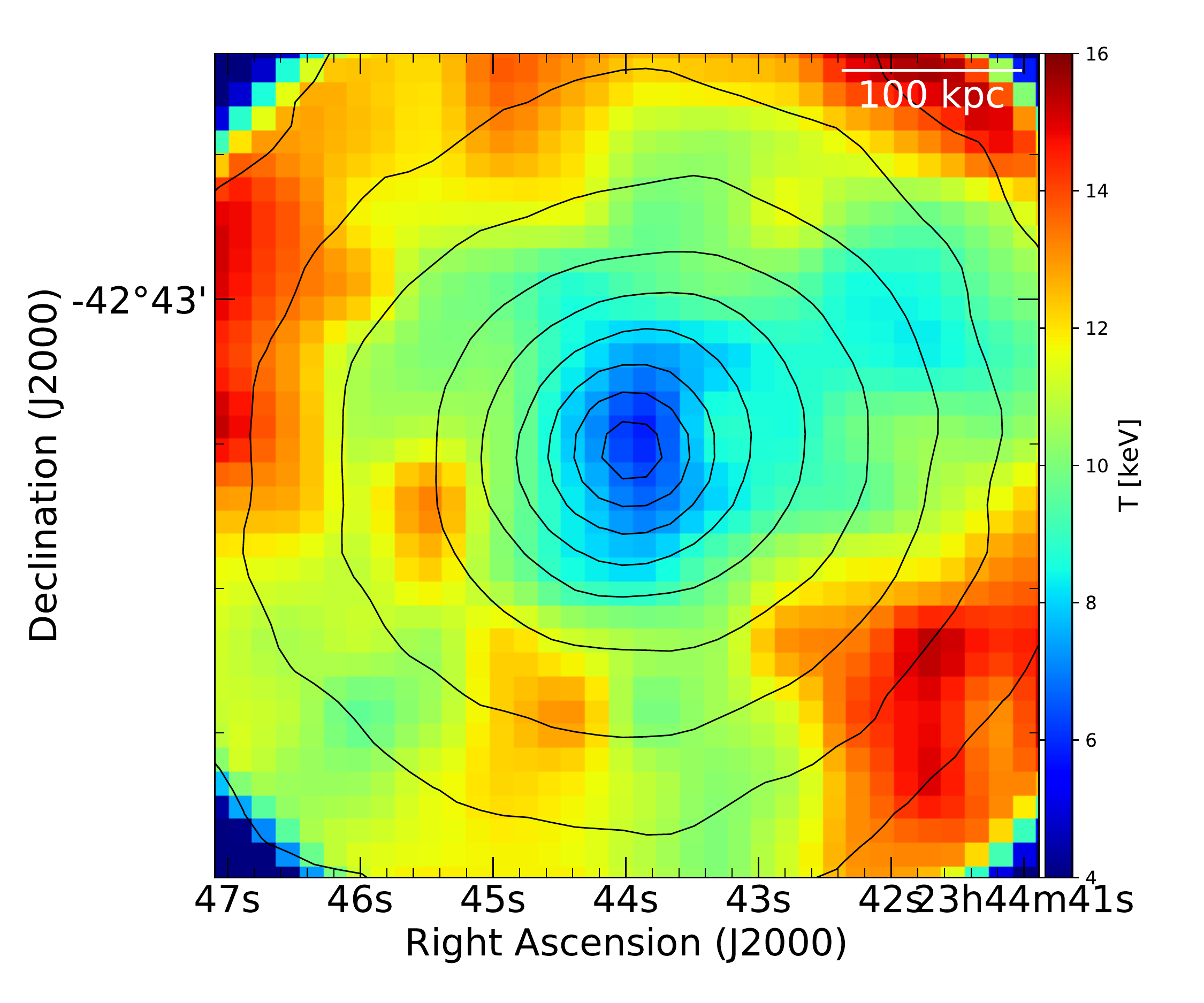} &
\includegraphics[width=\columnwidth]{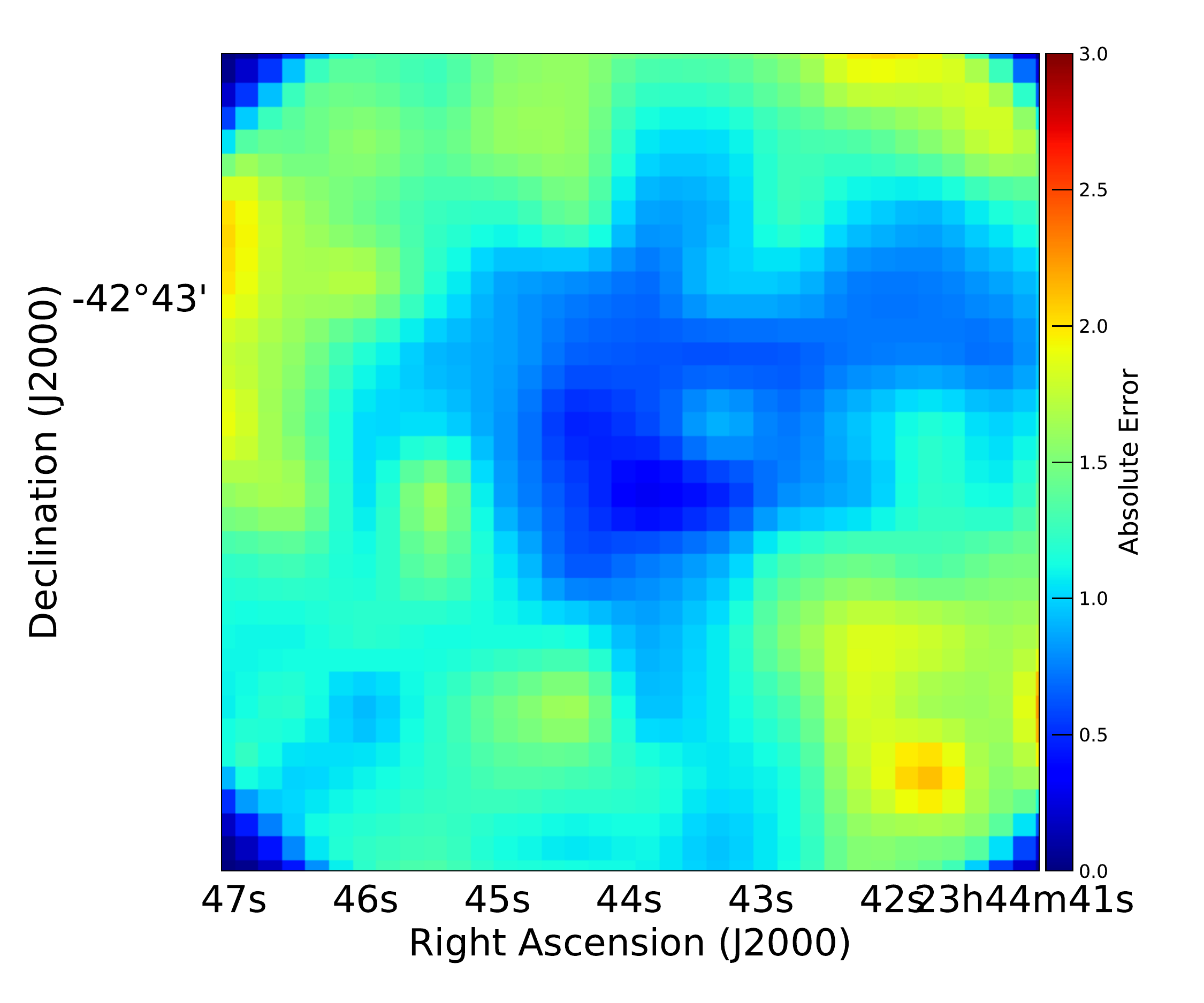}\\
\includegraphics[width=\columnwidth]{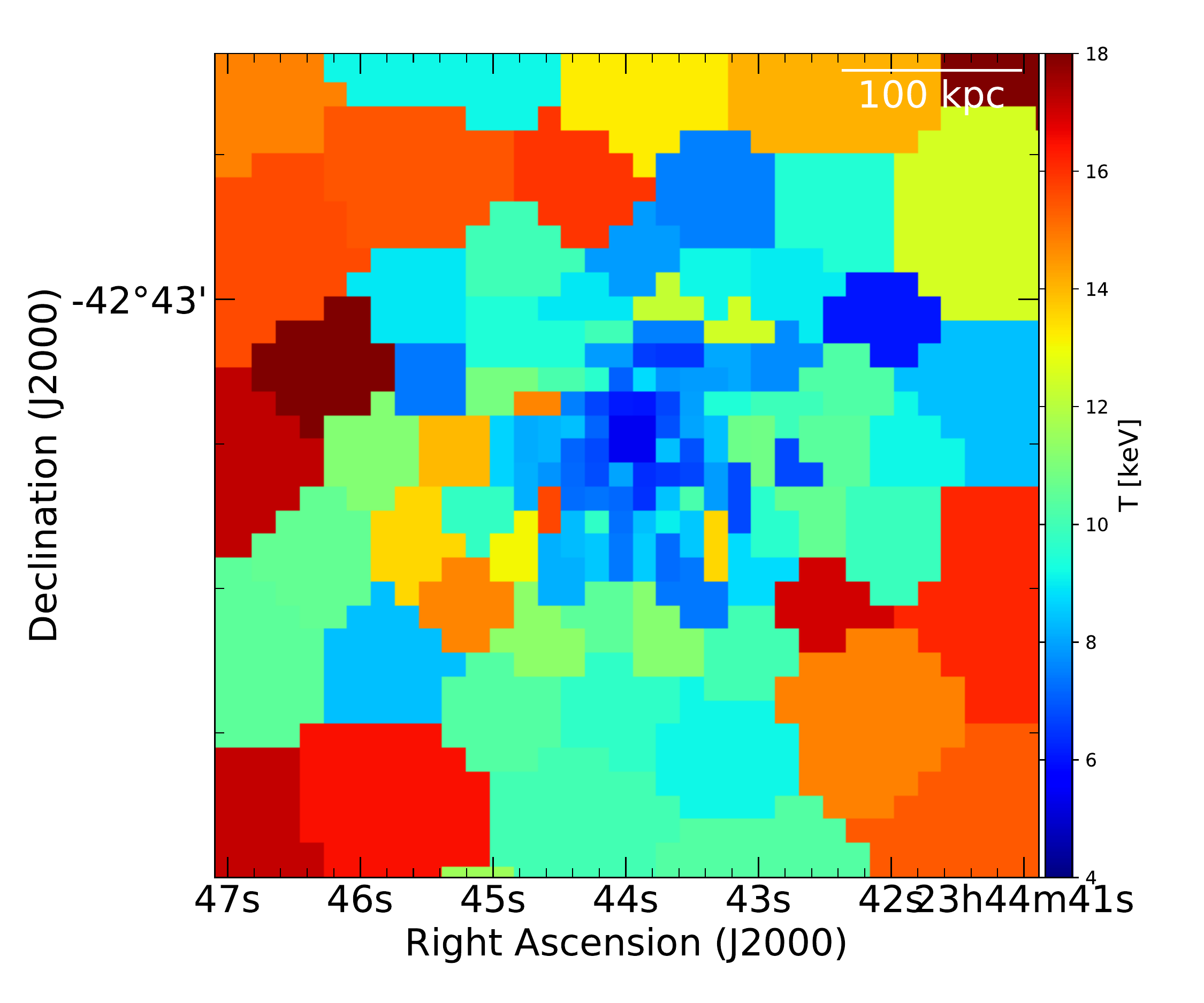} &
\includegraphics[width=\columnwidth]{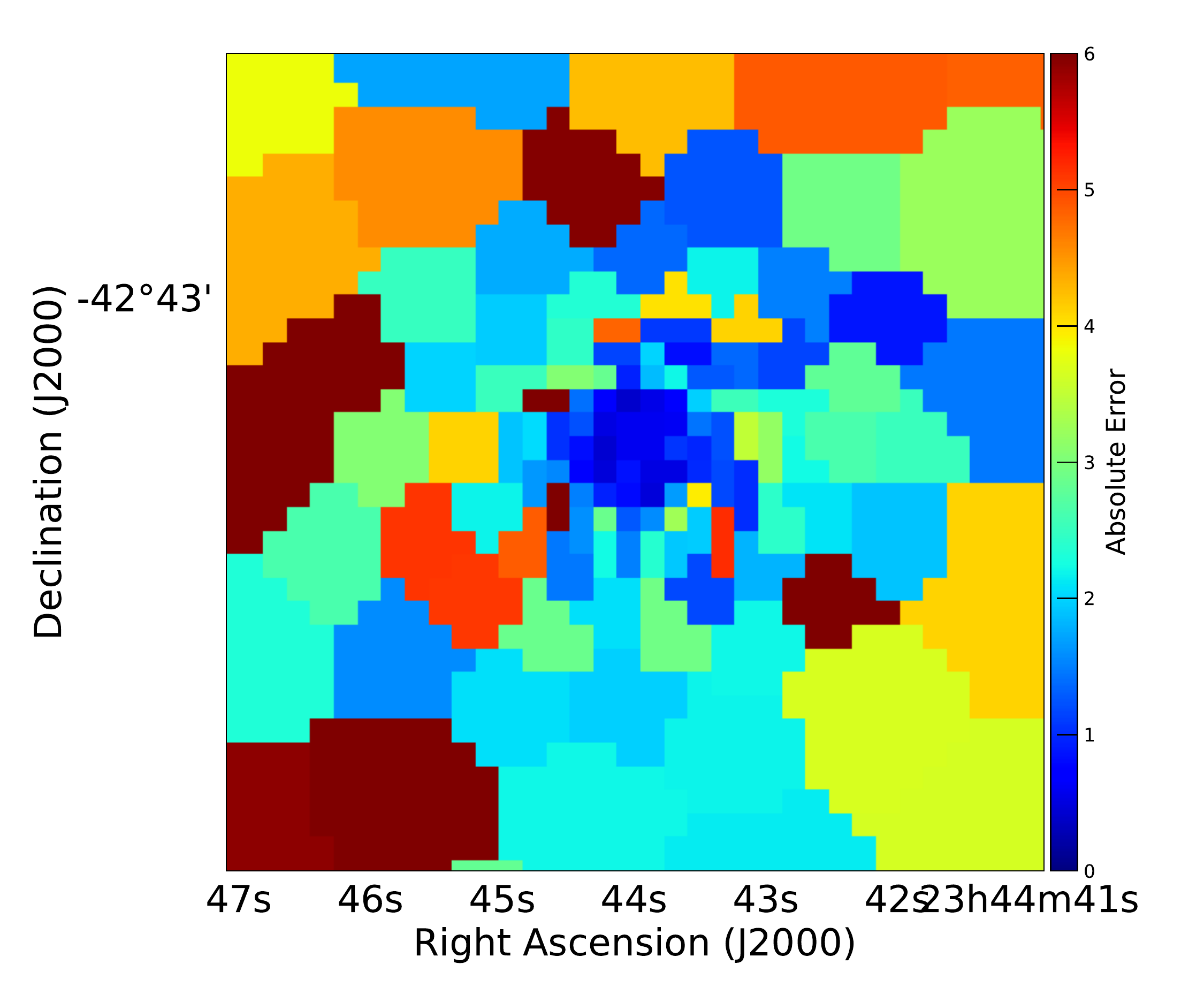}
\end{tabular}
\caption{{\it Top left:} $\it Chandra$ X-ray ACB temperature map (SN$=65$) overlaid with the same X-ray contours, as in Fig.~\ref{fig:xray}. {\it Top right:} Absolute error map associated with the temperature. Both of the images were smoothed across 3 pixels with a Gaussian kernel. The WVT temperature map (SN$=40$) (bottom left panel) and the temperature error map (bottom right panel) of the same.}
\label{fig:tmap}
\end{figure*}

This pipeline uses build-in \textit{CIAO} task\footnote{All CIAO tasks are documented here \url{http://cxc.harvard.edu/ciao/ahelp/ahelp.html}} \textit{download\_chandra\_obsid} to download data, then reprocess using \textit{chandra\_repro}. 
Because of the high redshift of the Phoenix cluster, there were available source free region in the ACIS-I chips during the observations. So, for background, we used this cluster emission free region of the ACIS-I chips that is $\sim$3 Mpc away from the cluster center.
To merge all the obsids, it uses \textit{merge\_obs} and \textit{fluximage} task. We used \textit{wavdetect} task to detect point sources in the 0.2-12 keV band, and checked visually whether the point sources were correctly chosen or not. After providing the SAO DS9\footnote{\url{http://ds9.si.edu/}} region file of the point sources, it excludes these point sources and creates a point source removed flux image (Fig. \ref{fig:xray}). High energy background flare time intervals were filtered out, and corresponding high energy wings were removed by generating light curves in the full band and 10-12 keV band for each observation. The light curves were binned at 259 sec per bin. Count rates greater than 3$\sigma$ from mean were removed using the task \textit{deflare}. We have visually inspected the light curve to ensure that the background flares were effectively removed. Spectra were extracted from circular regions in the ACB method which were created using method adapted from \citet{Randall2008,Randall2010}, and widely used by e.g., A2744 \citet{Owers2011}, A2443 \citet{Clarke2013}, A3667 \citet{Datta2014}, A85 \citet{Schenck2014}, A115 \citet{Hallman2018}. 
For WVT temperature map, we have used algorithm developed by \citet{Diehl2006} to create regions. For both methods, regions were just large enough to reach some minimum counts per spectra with applying threshold signal to noise ratio (SNR) as constrains. We have used SNR = 65 for ACB and SNR = 40 for WVT method to have more than 2000 counts per spectral region. Here, the signal is background subtracted counts, and noise is Poisson's noise coming from both signal and background. In ACB method, the best fitted temperature of each region was assigned to the pixel at the center of the circle, and every pixel has a corresponding circular spectral region. Circles are allowed to overlap so that some pixels will share counts from other pixels. Central 2.5$\arcsec$ region is contaminated by strong point source (AGN) \citep{Ueda2013,McDonald2019}. Following \citet{McDonald2019}, we masked this region and only considered 0.7-2.0 keV energy band for spectral fitting.
\textit{specextract} was used to derive the response files (RMF, ARF) and \textit{dmextract} for source and background spectra. 
We used XSPEC version 12.9.1 to fit these spectra within 0.7-7.0 keV energy range using the APEC thermal plasma model and the PHABS (photo-electric absorption) model. The best fitted results and their corresponding errors (using C-statistics; \citealt{Cash1979}) were then used to create the ACB (Adaptive Circular Binning) and WVT temperature maps (Fig. \ref{fig:tmap}). Most of these extraction routines were processed in parallel on a shared memory cluster computer with 96 cores and 512 GB RAM.

In the next section, we discuss the results from the observations starting with the diffuse radio emission.

\begin{deluxetable*}{lcccc}
\tablecaption{Integrated flux of the unresolved point sources \label{tab:pointFlux}}
\tablecolumns{5}
\tablewidth{0pt}

\tablehead{
\colhead{Source} & \colhead{RA DEC} & \colhead{$S_{610\ MHz}$} & \colhead{$S_{1.52\ GHz}$} & \colhead{$\alpha$} \\
\colhead{} & \colhead{(J2000)} & \colhead{(mJy)} & \colhead{(mJy)} & \colhead{}}
\startdata
A & 23 44 43.9 -42 43 12.5 & $72.42 \pm 0.08$ & $30.47 \pm 0.16$ & $-0.95 \pm 0.01$\\
B & 23 44 41.6 -42 43 21.7 & $3.66 \pm 0.08$ & $2.82 \pm 0.15$ & $-0.29 \pm 0.22$\\
C & 23 44 43.3 -42 43 35.3 & $3.15 \pm 0.12$ & $1.15 \pm 0.14$ & $-1.10 \pm 0.13$\\
D & 23 44 47.9 -42 43 42.6 & $3.50 \pm 0.13$ & $1.56 \pm 0.10$ & $-0.88 \pm 0.09$\\
E & 23 44 49.3 -42 42 54.0 & $1.07 \pm 0.08$ & $0.58 \pm 0.06$ & $-0.67 \pm 0.21$\\
\enddata
\end{deluxetable*}

\section{Diffuse radio emission from the cluster core} \label{sec:radio_emission}
The presence of diffuse radio emission in the Phoenix cluster was first detected by \citet{vanWeeren2014} at 610 MHz with GMRT observations and characterized as a minihalo. 
The central BCG was also detected in a 156 MHz GMRT observation \citep{vanWeeren2014} and in the 843 MHz Sydney University Molonglo Sky Survey \citep{Bock1999,Mauch2003} with flux densities of $330 \pm 45$ mJy and $79.2 \pm 2.8$ mJy, respectively.

\subsection{JVLA L-Band Data} \label{subsec: L Band diffuse}
In the left panel of Fig.~\ref{fig:radio_image-cont}, we present a 1.52 GHz radio image overlaid with 610 MHz contours. This is a low resolution image ($28\arcsec \times 17\arcsec$), which reveals the diffuse radio emission surrounding the central BCG. 
This image was made using a Briggs robust parameter of 0.5 and a Gaussian taper of $\sigma_{taper} = 7$ k$\lambda$ to emphasize large angular scale emission sensitivity.

The radio emission extends throughout most of the cluster region visible in the X-ray.
To estimate the amount of diffuse radio emission present in the cluster region, first we integrated the total emission enclosed within the $3\sigma_{\mathrm{rms}}$ contour of the low resolution image (Fig.~\ref{fig:radio_image-cont}, left panel). Then, a high resolution image was made using the robust parameter $-1$ and excluding baselines shorter than 7 k$\lambda$ (corresponding to a physical scale of about 240 kpc) to get rid of the large scale emission. The flux densities of the point sources were estimated using PyBDSF (Python Blob Detector and Source Finder; \citealt{Mohan2015}) with which we modeled these unresolved point sources using Gaussian fitting and estimated their corresponding integrated flux densities. These flux values are presented in Table \ref{tab:pointFlux}.
After that, the flux density contributions of these embedded point sources were subtracted from the total emission to get the integrated flux density of the diffuse radio emission. 
The minihalo emission at 1.52 GHz was found to be $S^{MH}_{\ 1.52\ GHz} = 9.65 \pm 0.97$ mJy.
The uncertainty in the flux density was estimated
using

\begin{equation}
    \sigma_{MH} = \sqrt{(\sigma_{cal}S_{MH})^2 + (\sigma_{\mathrm{rms}}\sqrt{N_{beam}})^2}\,,
    \label{eq:flux_err}
\end{equation}

\noindent where, $\sigma_{cal}$ and $\sigma_{\mathrm{rms}}$ are calibration uncertainty and map noise respectively. 
$N_{beam}$ is the number of beams present within the 3$\sigma_{\mathrm{rms}}$ contour. 
We assumed the $\sigma_{cal}$ to be $10\%$ in the flux density error estimation.

To estimate the size of the diffuse emission, we produced an unresolved point source subtracted image. This image is shown as contour overlay in Fig.~ \ref{fig:residual-background} (left panel).
Following \citet{Cassano2007},
we have estimated the radius of the diffuse radio emission using $R_{MH} = \sqrt{R_{max} \times R_{min}}$ where $R_{max}$ and $R_{min}$ are maximum and minimum radii measured corresponding to the 3$\sigma_{\mathrm{rms}}$ contour. This radius came out to be $R_{MH} = 0.31$ Mpc. 
The flux density of the minihalo within the 3$\sigma_{\mathrm{rms}}$ contour was estimated to be $11.0 \pm 1.05$ mJy, which is consistent within the errorbars with the previous estimate.
To verify the existence of the diffuse emission at 1.52 GHz, an alternative point source subtraction method is presented in Appendix Sect. \ref{sec:Apdx_resid_img}.

\subsection{GMRT 610 MHz Data}
In order to compute the spectral index between 610 MHz and 1.52 GHz, we made a 610 MHz image selecting a relevant UV-range such that the image resolution matches with that in the JVLA L band image. We used the same 7 k$\lambda$ UV-taper and a Briggs robust parameter of 0.5 as in the 1.52 GHz image. Finally, the image was restored with the same $28\arcsec \times 17\arcsec$ beam. This image is shown as white radio contours overlaid on the 1.52 GHz image in the left panel of Fig.~\ref{fig:radio_image-cont}. The extent of the radio emission in both frequencies is almost overlapping.

The minihalo emission at 610 MHz was estimated following the same procedure discussed above. The high resolution image was produced, excluding baselines shorter than 7 k$\lambda$ and setting the robust parameter to -1. The unresolved point sources were modeled using PyBDSF on this high resolution image (Table \ref{tab:pointFlux}). Finally, the flux density contribution corresponding to these point sources were subtracted from the total emission enclosed by the $3\sigma_{\mathrm{rms}}$ contour.
The minihalo emission at 610 MHz came out to be $S^{MH}_{\ 610\ MHz} = 22.54 \pm 2.26$ mJy. The flux density uncertainty was calculated using Eq.~\ref{eq:flux_err} assuming 10\% calibration error.

\subsection{Spectral index estimation}
We estimated the spectral index of the integrated minihalo emission between 610 MHz and 1.52 GHz by selecting the same region where radio emission is present in both frequencies within the $3\sigma_{\mathrm{rms}}$ contour. 
Then, we estimated the minihalo flux densities similarly to as described above. The spectral index between 610 MHz and 1.52 GHz was found to be $\alpha = -0.98 \pm 0.16$ which is on the flatter side of typical minihalo spectra reported earlier in the literature (e.g., \citealt{Sijbring1993, Giacintucci2011,Ferrari2011}). 
The uncertainty in the spectral index is estimated using 

\begin{equation}
    \Delta \alpha = \frac{1}{\log\Big(\frac{S_1}{S_2}\Big)}\sqrt{\Bigg(\frac{\Delta S_1}{S_1}\Bigg)^2 + \Bigg(\frac{\Delta S_2}{S_2}\Bigg)^2}\,,
    \label{eq:spec_index_err}
\end{equation}

\noindent where, $S$ and $\Delta S$ are flux density and respective uncertainty.
We calculated the k-corrected 1.4 GHz radio power of the minihalo using 

\begin{equation}
    P_{1.4\ GHz} = 4\pi S_{1.4GHz}\ D^{2}_L (1 + z)^{-(\alpha + 1)}\,,
    \label{eq:P1.4GHz_power}
\end{equation}

\noindent and was found to be $P_{1.4\ GHz} = (14.38 \pm 1.80) \times 10^{24}$ W Hz$^{-1}$ which is consistent with the X-ray luminosity and follows the observed $L_X-P_{1.4\ GHz}$ correlation \citep{Kale2015}.

\begin{figure*}[!t]
    \centering
    \begin{tabular}{cc}
    \includegraphics[width=\columnwidth]{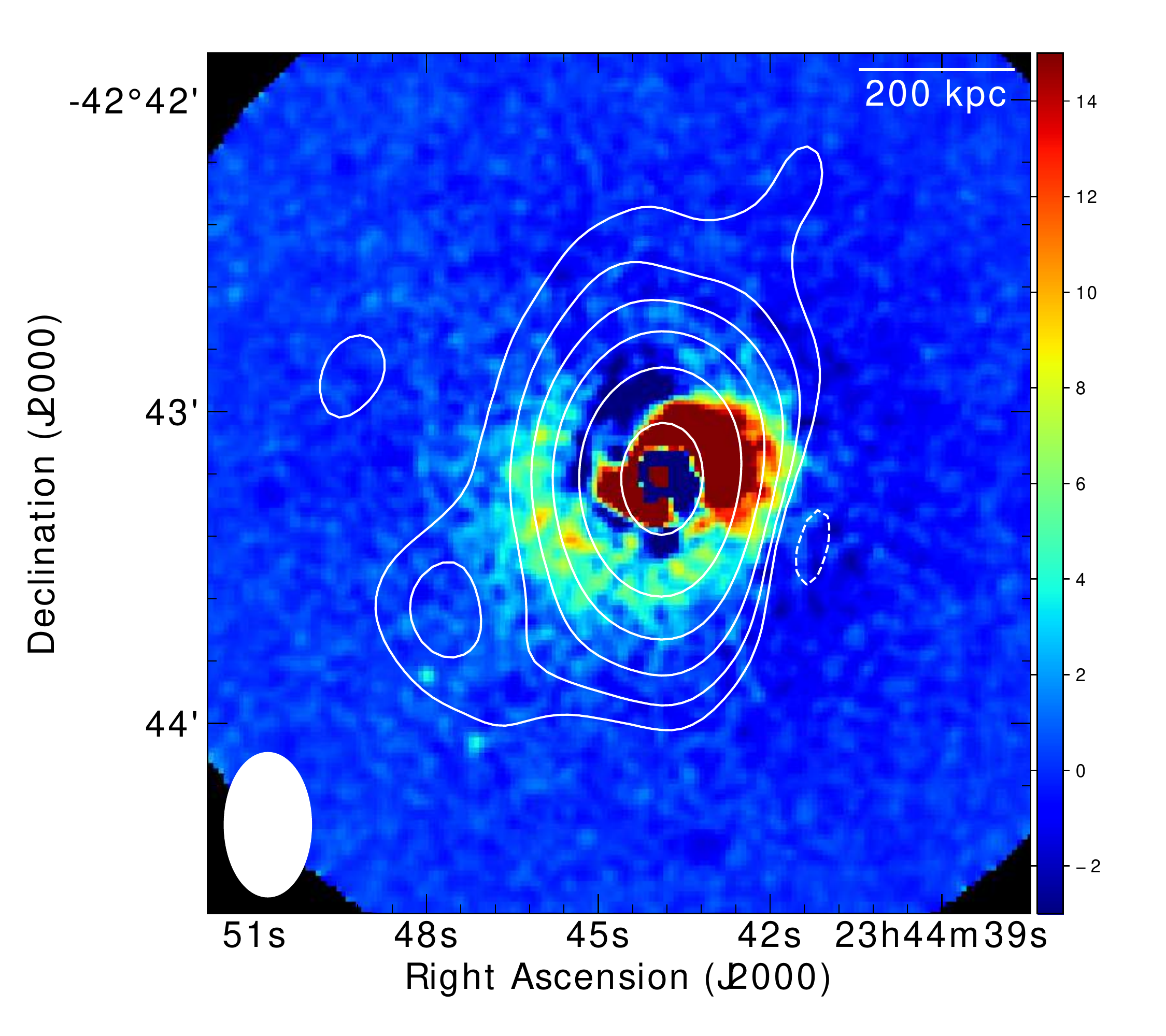} &
    \includegraphics[width=\columnwidth]{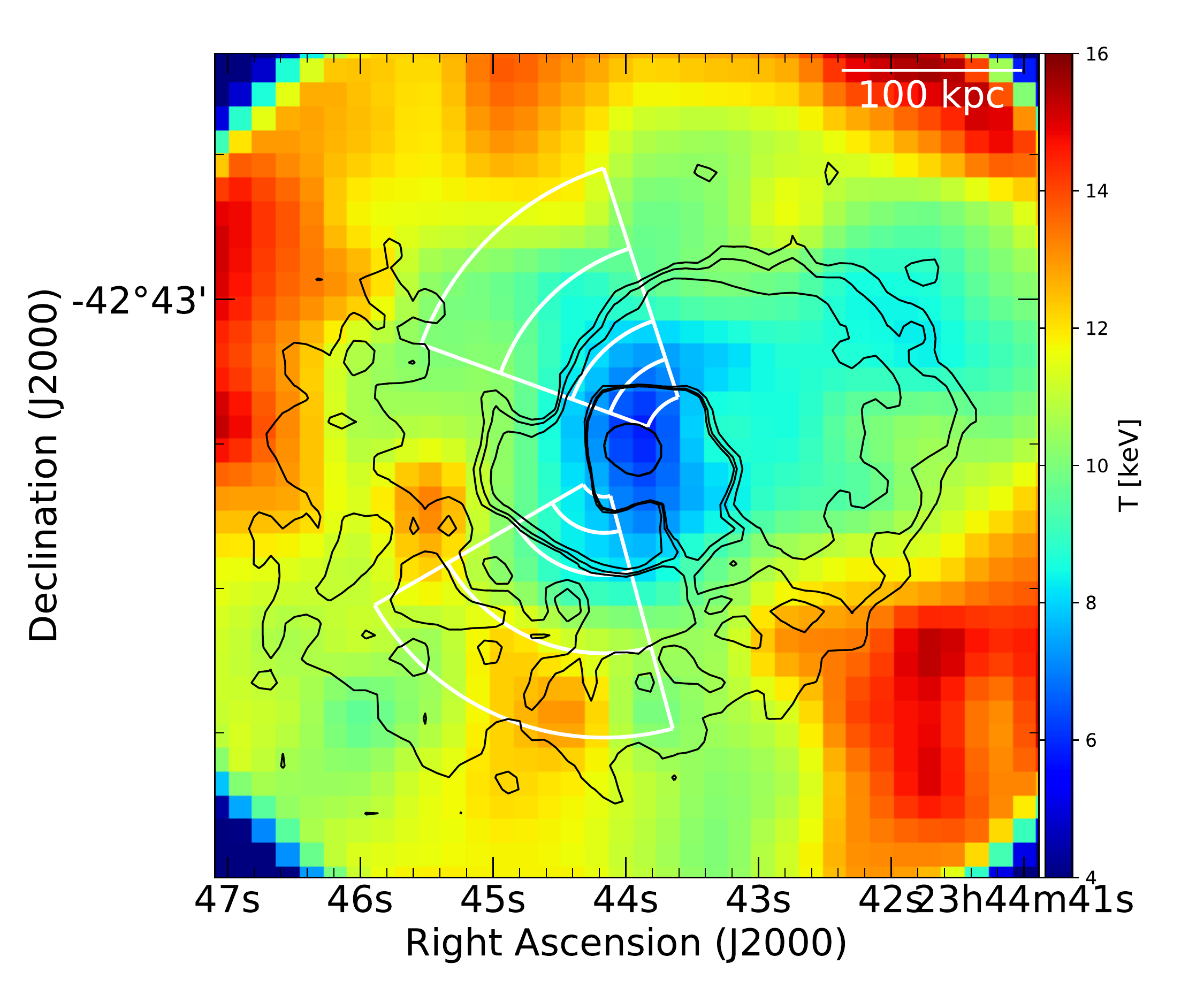}\\
    \end{tabular}
    \caption{{\it Left:} Smoothed X-ray brightness residual map after subtracting an elliptical $\beta$-model from the {\it Chandra} image. The unresolved point source subtracted 1.52 GHz radio image contours are shown in white at the same levels as in the right panel of Fig.~\ref{fig:radio_image-cont} but with $\sigma_{\mathrm{rms}}$= 30 $\mu$Jy beam$^{-1}$. {\it Right:} The temperature map overlaid with residual X-ray brightness contours in black. The contour levels are spaced by a factor of 2. The two sectors used for extracting the radial profiles are shown as white wedges.} 
    \label{fig:residual-background}
\end{figure*}

\section{Cluster Dynamics in X-ray} \label{sec:xray results}

\subsection{Temperature Map}
The cool core clusters are characterized by their regular morphology and extreme X-ray luminous core.
The Phoenix galaxy cluster is one of the most relaxed clusters with the strongest cooling flow known, possibly a rare example of runaway ICM cooling \citep{McDonald2019}. 
The X-ray image (Fig. \ref{fig:xray}) shows the dense, ultra-luminous core with near concentric brightness contours suggesting the absence of recent major merger events. 
On the other hand, the temperature maps shown in Fig.~\ref{fig:tmap} reveals the disturbance in the core region along with the thermal gas distribution of the ICM. 
These temperature maps are consistent with the previously published temperature map by \citet{McDonald2019a}.

Unlike the brightness map (Fig.~\ref{fig:xray}), the temperature distribution is not that regular. 
The left panel of Fig.~\ref{fig:tmap} clearly shows the cool core where the X-ray luminosity is highest. 
Hot gas is distributed surrounding most of the cool core, but the gas temperature on the eastern side of the core is much more uniform compared to the western counterpart.
Apart from the core, there is another region in the west where there is cold gas with almost similar temperature.
These two regions are connected with mildly cold gas with intermediate temperature.
There are also two hot gas clumps in the north and south of this cold region.
 
\subsection{Residual Brightness Map}
Fig.~\ref{fig:xray} shows that the X-ray surface brightness distribution from the cluster core towards the periphery is almost isotropic, which is also evident from the near concentric brightness contours. So, to explore more subtle changes in the brightness distribution, we produced a surface brightness unsharped mask map of this cluster (e.g., \citealt{Ichinohe2015}). To do that, we fitted a 2D elliptical $\beta-$model to the smoothed X-ray brightness image ($0.7-8.0$ keV).
Then, we subtracted the 2D $\beta-$model from the unsmoothed brightness image to obtain a residual brightness map. A smoothed residual brightness map is shown in the top left panel of Fig.~\ref{fig:residual-background} revealing excess surface brightness features that would otherwise be unseen in the original X-ray brightness map. 
The most prominent feature seen in this map is the spiraling brightness excess, where cool gas has formed a spiral shape.
The presence of spiraling cool gas in the Phoenix cluster was first reported by \citet{McDonald2015}, but the detection was of low significance because of the signal-to-noise limitation. Here in Fig.~\ref{fig:residual-background}, the over-dense region is clearly detected along with the eastern tail of the spiral which was barely detected by \citet{McDonald2015}.
It is present all around the core but is much brighter near the core than in the peripheral region. 
\citet{Ascasibar2006} investigated processes like this in numerical simulations and found that minor merger events in the cooling flow clusters can generate this kind of feature.
The spiral structure is remarkably similar to Fig. 3 in \citet{ZuHone2013} wherein a numerical simulation of a minor merger event has produced cold front.
A sudden drop in the excess brightness across the outer edge of the spiral is seen in the north-east (NE) and south-east (SE) directions, making an edge like feature, with the NE one being more extreme.
The top right panel of Fig.~\ref{fig:residual-background} displays the temperature map overlaid with the residual brightness map contours, showing that these edge-like features coincide with the positions in the temperature map where temperature gradients are observed.

\subsection{Sloshing Cold Fronts}
 
Even though the excess brightness is present in almost every direction from the cluster center, forming a spiral, sudden decreases are visible only in the NE and SE directions (Fig.~\ref{fig:residual-background}). 
For the quantitative measurement of these edges, we have extracted radial profiles along these directions, as indicated with white wedges in the top right panel of Fig.~\ref{fig:residual-background}. 
The NE wedge spans angles between 18 to 70 deg and the SE wedge angles between 120 to 195 deg where the position angles were measured from the north.

\begin{figure*}
    \centering
    \begin{tabular}{cc}
    \includegraphics[width=2.65in]{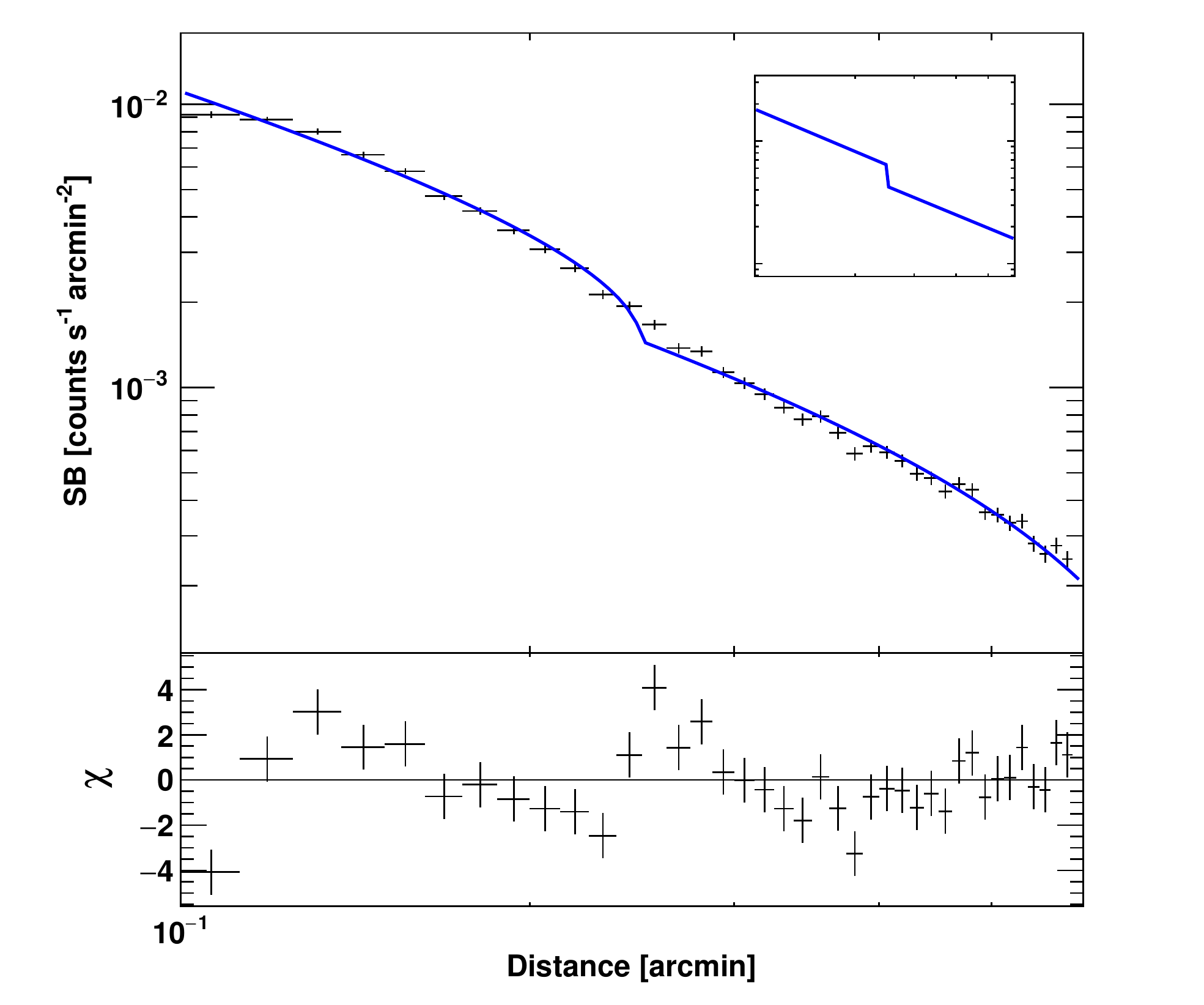} &
    \includegraphics[width=2.65in]{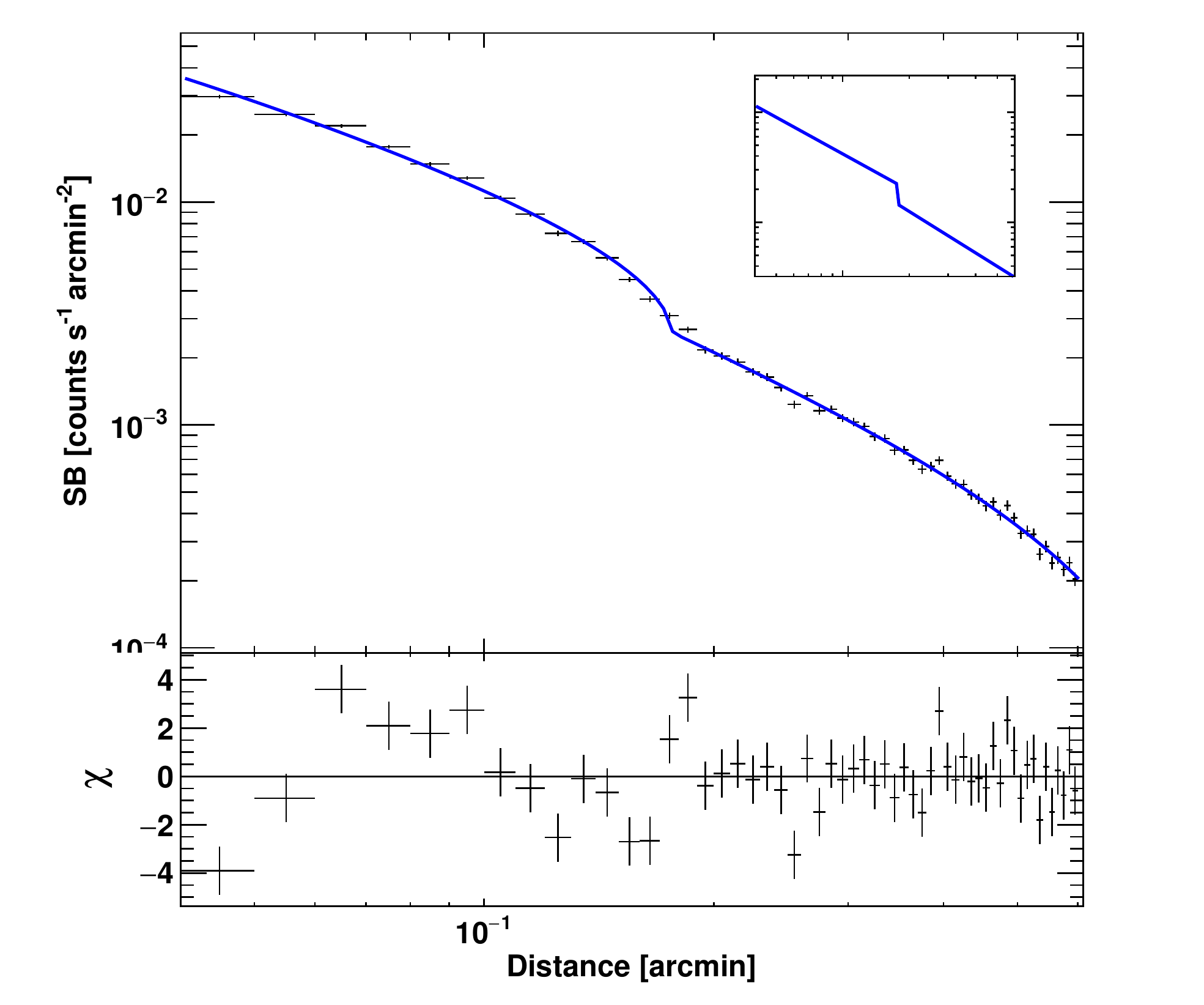}\\
    \includegraphics[width=2.7in]{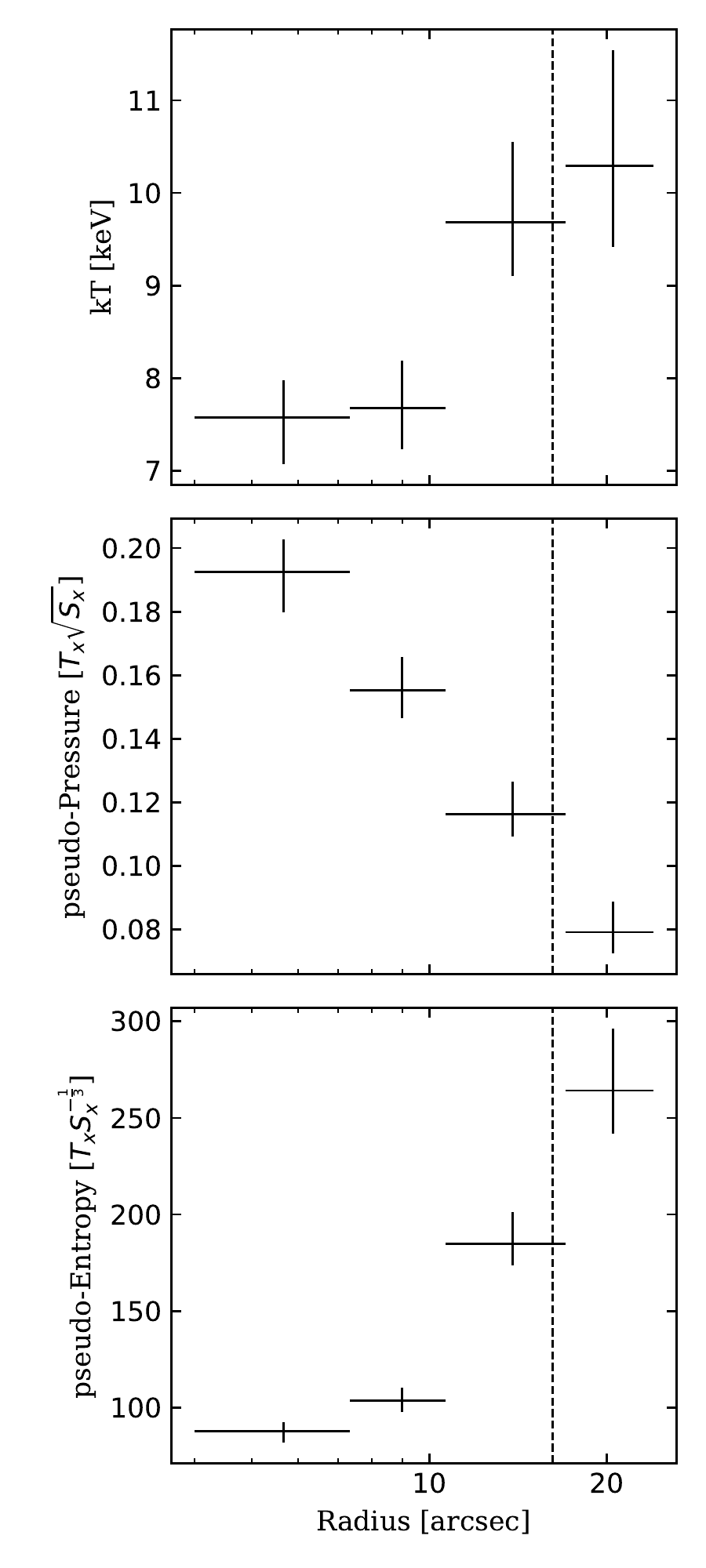} &
    \includegraphics[width=2.7in]{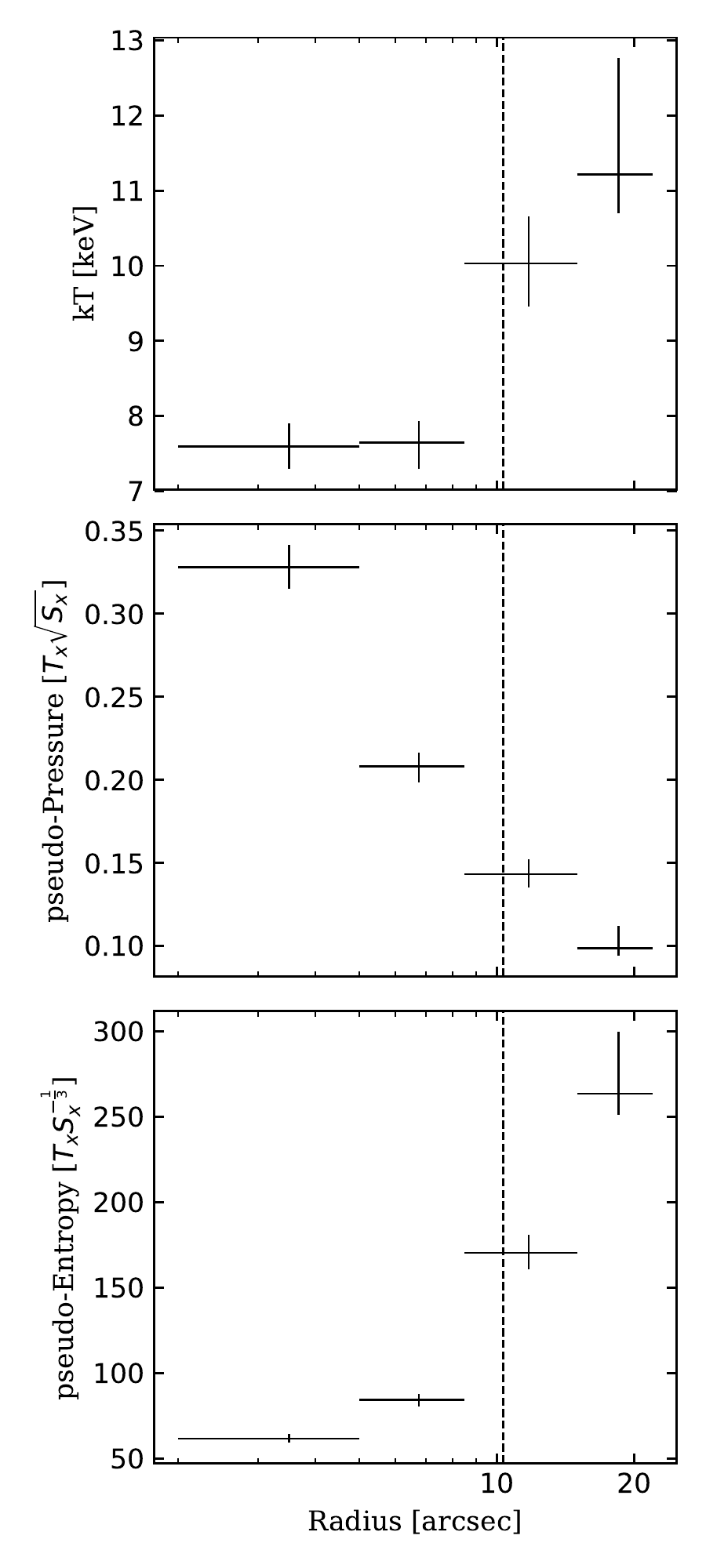}
    \end{tabular}
    \caption{Panels from top to bottom are X-ray surface brightness, projected temperature, pseudo-pressure, and pseudo-entropy profiles of the cluster regions corresponding to position angles $18-70$ deg (left panels) and $120-195$ deg (right panels) as indicated in Fig.~\ref{fig:residual-background} top right panel. The position of the density discontinuity is indicated with the vertical dashed line.} 
    \label{fig:jumps}
\end{figure*}

In Fig.~\ref{fig:jumps}, we present X-ray surface brightness, projected temperature, pressure, and entropy profiles for these wedge regions. 
The surface brightness profile, as well as the broken powerlaw fit was performed with PROFFIT developed by \citet{Eckert2011}.
We have extracted the temperature profile by fitting the X-ray data corresponding to each annular similar to as done in the ACB temperature map generation. 
Combining the mean brightness corresponding to each annular and the extracted temperature, we derived pseudo-pressure and pseudo-entropy profiles as well. 

As seen in the unsharped mask map, temperature discontinuity is seen across the NE and SE edges. The corresponding X-ray brightness shows relatively subtle change across these edges,
and also the pressure varies smoothly. 
The top panels of Fig.~\ref{fig:jumps} clearly shows the presence of density discontinuity across these edges. The density jump across the NE and SE edges are $1.33 \pm 0.03$ and $1.37 \pm 0.04$ with reduced $\chi^2$ being 1.37 and 1.02, respectively.
These are cold fronts or contact discontinuities, which are ubiquitous in cool core clusters, indicating gas sloshing in the cluster core. 
These sloshing cold fronts are possibly induced by minor merger events \citep{Ascasibar2006}.
The positions of the NE and SE cold fronts are at $\sim$98 and $\sim$69 kpc, respectively, from the cluster center, taken to be the X-ray brightness peak.
For a systematic investigation, we have searched for projected temperature and density discontinuity in all directions following the X-ray excess brightness spiral and found none other than those mentioned above.

\section{Summary and Conclusions} \label{sec:conclude}
Our high resolution {\it Chandra} temperature map presents a cluster-wide thermal gas distribution, and the unsharped mask map reveals the spiraling cool gas around the cluster core resulting from a past minor merger event.
Two subtle cold fronts in the NE and SE of the core are detected, which indicate the presence of sloshing motion in the cluster. These are possibly the most distant resolved cold fronts detected to date.

The diffuse radio emission discovered in the 610 MHz observations is also present at 1.52 GHz, occupying a similar cluster region with a radius of about 0.31 Mpc.
The minihalo emission at 1.52 GHz was found to be $6.65 \pm 0.97$ mJy.
We calculated the spectral index of the minihalo between 610 MHz and 1.52 GHz to be $-0.98 \pm 0.16$, which is flatter than the typical minihalo spectrum.
The top left panel of Fig.~\ref{fig:residual-background} shows that where the cold fronts are found in the vicinity of the cluster core, diffuse radio emission is present on a much larger scale.
Other such examples were reported by, e.g.,~\citet{Giacintucci2011,Giacintucci2014,Savini2018,Savini2019}, where minihalo emission is present well beyond the cold fronts. 
Fig.~\ref{fig:residual-background} also shows the large scale spiraling cool gas, indicating that the gas sloshing is happening on a much larger scale, which covers most of the minihalo region.
So, even if the large scale cold fronts have decayed, the turbulence is still present, providing the necessary energy to the {\it in situ} relativistic electrons.

It is worth pointing out that \citet{McDonald2019a} reported the presence of radio jets from the central BCG in the Phoenix cluster, but our 1.52 GHz observation lacks the necessary spatial resolution to quantify the radio emission from these jets accurately. So, the possibility of jet emission contamination in the flux density estimation at 1.52 GHz, resulting in the relatively flatter spectrum, can not be ruled out. In addition, Fig. \ref{fig:residual-background} shows that surprisingly, the minihalo emission extends in the direction of the BCG jets (Fig. 5 \citealt{McDonald2019a}) further indicating the possible AGN influence in the ICM.
Since the possible contamination by the AGN jets can not be ruled out, the estimated spectral index value should be regarded as an upper limit to the minihalo spectral index.


In conclusion, the sloshing driven turbulence seems to be playing a major role in providing the necessary relativistic electrons for minihalo emission. 
Future low frequency radio observations with higher large angular scale sensitivity are needed to clarify the true extent of the minihalo emission and also to look for the presence of steep spectrum large scale emission.
Also, future high frequency radio observations with good large scale sensitivity are needed to detect the presence of spectral break, if any, which would be a signature of the re-acceleration model. 
Most importantly, resolved multi-frequency observations are necessary to quantify the BCG and jet emission contributions accurately.

\acknowledgments
We thank IIT Indore for providing computing facilities necessary for the data analysis.~We thank both the VLA and GMRT staff who made these radio observations possible. GMRT is run by the National Centre for Radio Astrophysics of the Tata Institute of Fundamental Research.~RR is supported through ECR/2017/001296 grant awarded to AD by DST-SERB, India. MR would like to thank DST-India for INSPIRE fellowship program for financial support (IF160343). This research was supported in part by NASA ADAP grant NNX15AE17G. DR is supported by a NASA Postdoctoral Program Senior Fellowship at the NASA Ames Research Center, administered by the Universities Space Research Association under contract with NASA.
\facilities{JVLA, GMRT, {\it CXO}}
\software{CASA \citep{McMullin2007}, SPAM \citep{Intema2009,Intema2017}, XSPEC \citep{Arnaud1996}, PyBDSF \citep{Mohan2015}, ClusterPyXT \citep{Alden2019}, CIAO \citep{Fruscione2006}, PROFFIT \citep{Eckert2011}}


\appendix
\section{Alternative point source subtraction method in the 1.52 GHz image} \label{sec:Apdx_resid_img}
To verify the existence of diffuse radio emission at 1.52 GHz, we have presented the direct point source subtracted 1.52 GHz image of the Phoenix cluster (unlike the contours presented in Fig. \ref{fig:residual-background} left panel where the point source contributions were subtracted from the UV-data). Here, first, we have convolved the high resolution point source image (Fig. \ref{fig:radio_image-cont} right panel) with the restoring beam of the low resolution image shown in Fig. \ref{fig:radio_image-cont} left panel. Then, we subtracted the new low resolution point source image from Fig. \ref{fig:radio_image-cont} left panel image. The resultant residual image is similar to the contour image presented in Fig. \ref{fig:residual-background} left panel. The integrated flux of this residual diffuse emission is also consistent with reported values (Sect. \ref{subsec: L Band diffuse}) within the errorbars.

\begin{figure}
    \centering
    \includegraphics[width=6.8in]{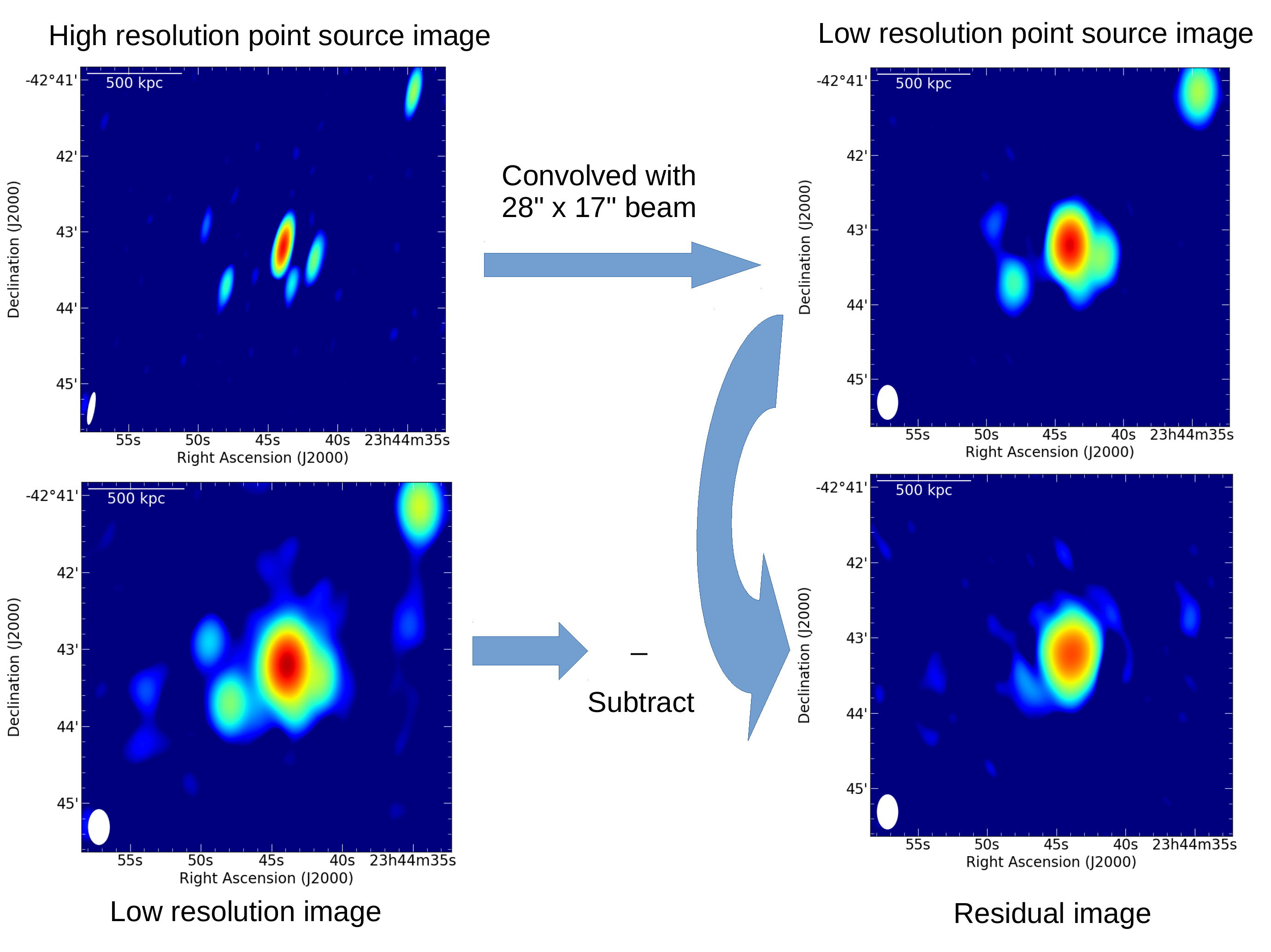}
    \caption{Direct point sources subtracted residual image of the diffuse emission at 1.52 GHz in the Phoenix cluster.}
    \label{fig:Apdx_resid_img}
\end{figure}

\end{document}